\documentclass[twocolumn,showpacs,preprintnumbers,amsmath,amssymb]{revtex4}

\usepackage{graphicx}
\usepackage{dcolumn}
\usepackage{bm}

\begin{document}


\title{Microscopic Wrinkles on Supported Surfactant Monolayers}

\author{Quan Zhang}
\email{quanz@uchicago.edu}
\author{T.A.Witten}%
\email{t-witten@uchicago.edu}
\affiliation{%
Department of Physics and James Franck Institute, University of
Chicago, Chicago, IL, 60637 USA\\ }%

\date{\today}

\begin{abstract}
We discuss mechanical buckling instabilities of a rigid film under
compression interacting repulsively with a substrate through a thin
fluid layer. The buckling occurs at a characteristic wavelength that
increases as the 1/4th power of the bending stiffness, like a
gravitational instability studied previously by Milner et al.
However, the potential can affect the characteristic buckling
wavelength strongly, as predicted by Huang and Suo. If the potential
changes sufficiently sharply with thickness, this instability is
continuous, with an amplitude varying as the square root of
overpressure.  We discuss three forms of interaction important for
the case of Langmuir monolayers transferred to a substrate:
Casimir-van der Waals interaction, screened charged double-layer
interaction and the Sharma potential. We verify these predictions
numerically in the Van der Waals case.
\end{abstract}

\pacs{46.32.+x, 46.70, 64.70}
\maketitle

\section{INTRODUCTION}

The advent of controlled molecular scale films and deposition
methods has revealed a number of fine-scale wrinkling instabilities
\cite{WhitsidesMetalFilm, SoftMatterHardSkin,
Ortiz.folding.patterns}.  At the same time, several new and general
features in the buckling of macroscopic films have been identified
\cite{Audoly.blisters, Cerda.wrinkling, Pomeau.Rica}.  Some of these
are elaborations of the simple Euler buckling of a compressed rod or
sheet \cite{ Love, Euler}. Recently Cerda and Pocivavsek \cite{luka}
have considered rigid, compressed sheets on the surface of a liquid
in the presence of gravity.  This extends an earlier treatment of
Milner, Joanny and Pincus \cite{mi-89} adapted to lipid monolayers
at an air-water interface.  Under these conditions the Euler
buckling occurs not at zero wavevector but at a finite wavevector
determined by the bending modulus and liquid density.  We call this
mode of buckling gravity-bending buckling.

Folding structure of laterally-compressed surfactant monolayers at
the air-water interface is a well-known phenomenon \cite{Kayee_1996,
Kayee_1997}. The initial instability leading to these folds may be
related to the gravity-bending buckling noted above.  The observed
folding length scale resembles the predicted wavelength of
gravity-bending buckling \cite{luka}. Analogous folding has recently
been observed in solid nanocrystal monolayers \cite{klara_2007}.

Additional topographic structure is observed when compressed lipid
monolayers are transferred to a solid substrate via the inverted
Langmuir-Schaeffer method \cite{Kayee_1998}.  These supported
monolayers and bilayers are increasingly common in the study of
biological membranes \cite{Kinlok_2006, Sackmann_2005, Guohui.Wu,
Muresan_2001,Sackmann_1996}.  The layer thus transferred is
positioned for easier study.  Initially these transferred layers are
separated from the substrate by a cushion of the carrier liquid. Any
topographic patterning of these transferred layers can readily be
observed \cite{Guohui.Wu, Shelli_2004}.  Such patterning is likely
affected by interaction with the substrate. Likewise, any buckling
of a supported monolayer must be affected by the substrate. To study
transferred monolayers under the high compressions where buckling is
expected seems feasible, though to our knowledge no such studies
have been performed.

In this paper we investigate a class of wrinkling instabilities that
are generalizations of gravity-bending buckling. These instabilities
occur when a deformable surface such as a lipid monolayer lies above
a solid substrate on a cushion of fluid.  The interaction of the
surface with the substrate can then play the role of gravity.  This
interaction alters the gravity-bending instability in several ways.
It makes the unstable wavelengths depend on depth $d$. These
wavelengths generally are much smaller than those predicted by the
gravity-bending instability.

In order to investigate the effect of the substrate in its simplest
form, our treatment neglects several effects that may be important
in practice \cite{Hobart_2000}. In practice some external forcing on
the film is required in order to create the lateral pressure to
induce buckling. For example regions outside the region under study
may be bonded to the substrate. In practice the time dependence of
the buckling may be important in determining its wavelength. This is
especially true in cases where the equilibrium buckling transition
is discontinuous. These effects have been extensively explored in
the semiconductor film literature \cite{liang_2002, Huang_2002,
Sridhar_2002}. Below we shall merely assume that a uniform unixial
stress is imposed on the region in question and will only
investigate the initial buckling instability as influenced by the
substrate interaction. Further, we shall suppose that the buckling
film is inextensible. In the appendix we show that the inextensible
approximation is appropriate for the lipid monolayers such as those
of Ref. 11 and 12.

Moreover, the interaction alters the qualitative nature of the
instability.  The gravity-bending instability is a runaway or
subcritical instability at constant surface pressure.   Here the
amplitude of the wrinkles jumps from zero to a large value
determined by other aspects of the system. However, substrate
interactions can change this behavior, making the amplitude a
continuous function of surface pressure.  The criterion for a
continuous transition can be stated generally in terms of the second
and fourth derivative of the interaction potential with separation.

In the following sections we discuss the substrate-bending
instability in terms of a general interaction potential $\phi(d)$.
We determine the unstable wavelength and the condition for a
continuous transition.  In the following section we consider three
specific potentials commonly encountered in liquid films.  The first
is the Casimir-van der Waals interaction: $\phi(d) = A/(12\pi d^2)$.
The second is the screened charged double-layer interaction common
in aqueous films with charged interfaces.  The third is the Sharma
potential used to describe molecularly thin water films.  All of
these potentials produce continuous wrinkling for sufficiently thick
films. We conclude that this substrate-bending buckling should be
readily observable.

\section{THEORY OF WRINKLING INSTABILITY}

\subsection{Wavelength of Microscopic Wrinkles}

In the analysis below, we will consider a simplified model of a
Langmuir monolayer. Suppose an insoluble surfactant layer is sitting
at the interface between air and a liquid subphase. The elastic
property of a layer, with finite thickness $t$, is characterized by
the bending modulus B \cite{lan-86}:

\begin{equation}
B = \frac{E t^3}{12(1-\nu^2)} \, ,
\end{equation}
where $E$ is the Young's modulus and $\nu$ is the Poisson ratio in
the continuum theory. A solid substrate is placed under the
subphase, as shown in Fig.\ref{fig:2:flatbuckled}. Interaction
between the solid substrate and liquid subphase takes the form of a
substrate potential $\phi(d)$ \cite{de-03}. For different
interactions, the functional forms of $\phi(d)$ are different
\cite{is-85}. For example, if the interaction is of pure van der
Waals type, we have:

\begin{figure}[!htp]
\begin{center}
\includegraphics[angle=-90, width=3.8in]{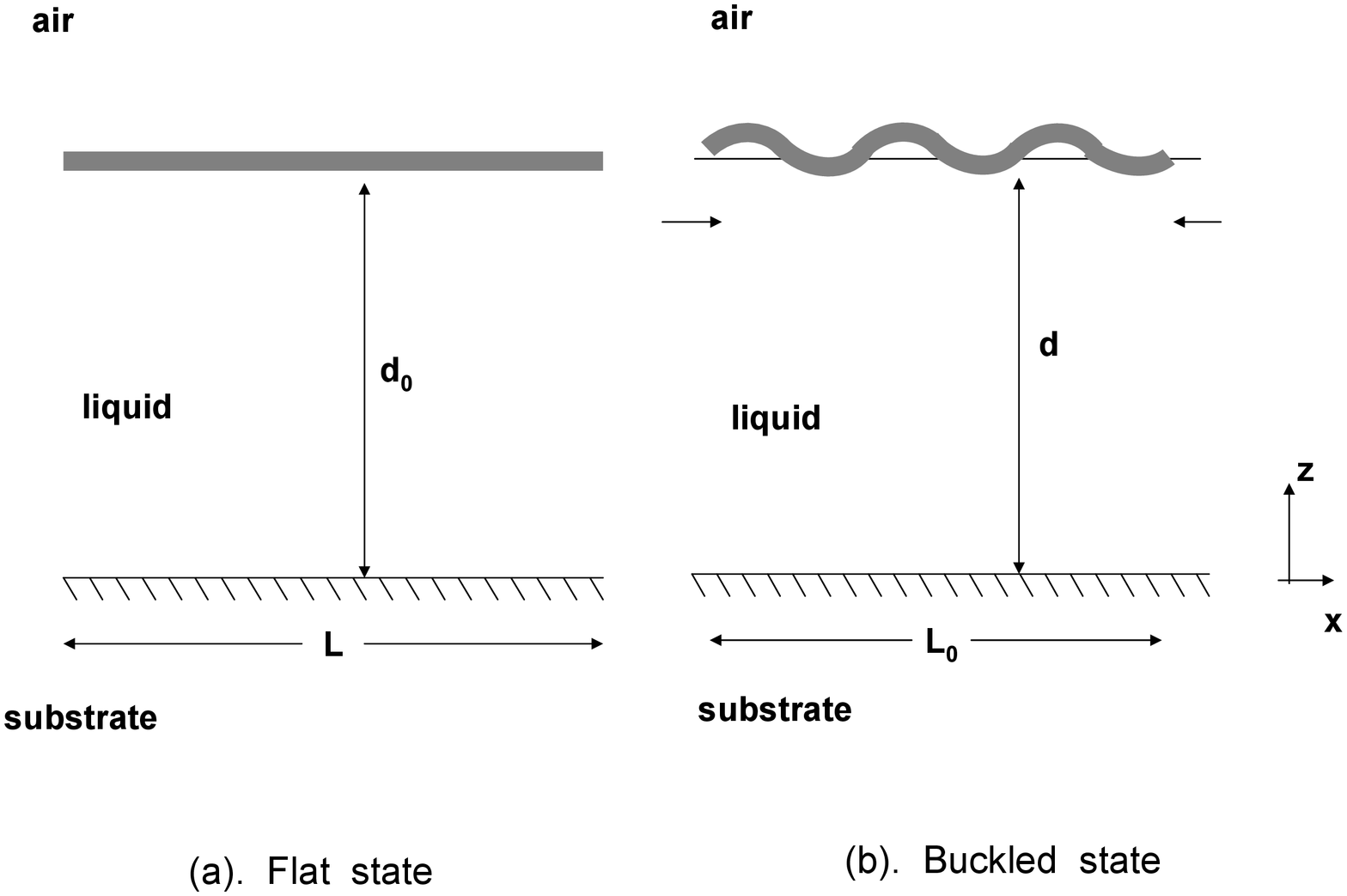}
\caption{One dimensional model used in the theory. The y-axis is
pointing out of the figure. (a). Initial Flat state with no
compression. (b). Buckled state with large enough compression.}
\label{fig:2:flatbuckled}
\end{center}
\end{figure}

\begin{equation}
\phi(d) = \frac{A}{12\pi d^2}  \, ,
\end{equation}
where $A$ is the Hamaker constant \cite{de-03}. It can take positive
or negative values depending on properties of substrate and
subphase. Possible retardation effects are not considered in this
paper. In a Langmuir trough, one can compress the surfactant layer
with external pressure $\Pi_{ex}$. Elastic strain energy is stored
in the elastic layer upon compression. It is expected that if
$\Pi_{ex}$ exceeds some critical value $\Pi_c$, the elastic layer
will enter a buckled state and relax the strain energy in a third
dimension, similar to the Euler buckling of a rod. We recall that
the critical pressure $\Pi_c$ is independent of the compressibility
of the layer, although the corresponding strain depends on the
compressibility \cite{Love}. In the following analysis a constant
external pressure $\Pi_{ex}$ is exerted on the Langmuir monolayer. A
buckling transition is induced by displacing the boundary. In the
wrinkled state, total area $S$ of the surfactant layer is:

\begin{equation}
S = \int_{S_0} \left(1 + (\nabla \xi)^2\right)^{1/2} \, ds  \, .
\label{eqn:w:inex}
\end{equation}
The above integral is taken over the projected area $S_0$ on the
horizontal x-y plane and $ds$ is a surface element in $S_0$. The
quantity $\xi(x,y)$ is the vertical displacement of interface from a
flat state. The height profile of the interface is: $d(x,y)= d_0 +
\xi(x,y)$, where $d_0$ is the height of a flat state with no surface
deformation. We assume that no subphase fluid enters or leaves, so
that the volume of subphase under the initial flat surface is fixed
during deformation:

\begin{equation}
\int_{S_0} \xi (x,y) \,  ds = 0  \, . \label{eqn:w:vol}
\end{equation}

Another parameter needed is the surface density of surfactant:
$\sigma = N/S$, where $N$ is the total number of surfactant
molecules. For constant external pressure $\Pi_{ex}$, the Gibbs free
energy is written as \cite{mi-89,tw-20}:

\begin{equation}
G = \gamma_0(S-S_0)+ F_{l}+ F_{b}+F_{i}+ \Pi_{ex} S_0  \, ,
\label{eqn:w:free}
\end{equation}
where $\gamma_0$ is the surface tension of a free interface without
any compression. The first term denotes the change in interfacial
energy. $F_{l}$ is the surfactant free energy. It is related to
$\sigma$ via the relation: $F_{l} = S f_l(\sigma)$. The last term is
an analogy of the pressure $P$ times volume $V$ term in the Gibbs
free energy of a conventional gas. $F_{b}$ and $F_{i}$ are bending
energy and substrate potential energy, respectively. For small
surface deformation ($|\xi/d_0|\ll 1$), the bending energy $F_b$ is
given by the Helfrich energy \cite{Helfrich_73,Hu-96}:

\begin{eqnarray}
F_b &=& \frac{B}{2}\int_{\textrm{material surface}} (C-C_0)^2 \nonumber \\
&=& \frac{B}{2}\int_{S_0}(C-C_0)^2 \left(1 + (\nabla
\xi)^2\right)^{1/2} \, ds  \, ,
\end{eqnarray}
where $C(x,y)$ is the mean curvature of the interface and $C_0$ is
the spontaneous curvature of surfactant layer. Again we have
expressed the integral in terms of the projected area as in equation
(\ref{eqn:w:inex}). $C_0$ has no effect on the following analysis
and we shall neglect it henceforth \cite{tw-20}. The substrate
potential energy is:

\begin{equation}
F_i = \int_{S_0} \phi(d) \left(1 + (\nabla \xi)^2\right)^{1/2} \, ds
- \phi(d_0)S  \, ,
\end{equation}
where the initial flat state is chosen as reference state for the
potential energy. Our treatment can be simplified by supposing that
the film is {\it inextensible}, as justified in the appendix. In
that case, the molecular density $\sigma$ is fixed and the energy
$F_l$ is a mere constant. Then for different values of external
pressure $\Pi_{ex}$, the equilibrium configuration of the system is
obtained by minimizing the Gibbs free energy with the
inextensibility constraint of the surfactant layer: $S = const$.
Introducing a Lagrange multiplier $\theta$ to incorporate this
constraint, the functional that we need to minimize is:

\begin{equation}
G' = G - \theta \left(S - \int_{S_0} (1 + (\nabla \xi)^2)^{1/2}\, ds
\right) \, .
\end{equation}

In the rest of this paper, we assume that relaxation of strain
energy only occurs in the x-direction, as shown in
Fig.\ref{fig:2:flatbuckled}. In the y-direction, the system has
translational invariance. As a result, we may minimize the Gibbs
free energy per unit length in y-direction, which we denote as $g'$:

\begin{eqnarray}
g' &=& g - \theta \left(L - \int_{L_0} (1 + (\dot{\xi})^2)^{1/2}
\, dx \right) \nonumber \\
&=& \gamma_0(L-L_0)+ L f_l(\sigma)+ f_{b}+f_{i}+ \Pi_{ex} L_0
\nonumber \\
&& - \theta \left(L - \int_{L_0} (1 + (\dot{\xi})^2)^{1/2}\, dx
\right) \, , \label{eqn:w:g'}
\end{eqnarray}
where $L$ is the total length of surfactant layer and $L_0$ is the
projected length in x-direction, as shown in
Fig.\ref{fig:2:flatbuckled}. In the above expression, $\dot{\xi}$
denotes the derivative of $\xi$ with respect to the x-coordinate. In
the one dimensional model, the mean curvature $C$ is given by:

\begin{equation}
C = \ddot{\xi}\cdot\left(1+\dot{\xi}^2\right)^{-3/2}  \, .
\end{equation}
Using this expression the bending free energy(per unit length in
y-direction) $f_{b}$ and its expansion in small deformation
approximation take the form:

\begin{eqnarray}
f_b &=& \frac{B}{2}\int_{L_0} (\ddot{\xi})^2
(1+\dot{\xi}^2)^{-5/2}\, dx
\nonumber \\
&\sim& \frac{B}{2}\int_{L_0}(\ddot{\xi})^2
\left(1-\frac{5}{2}\dot{\xi}^2+\frac{35}{8}\dot{\xi}^4
-\cdots\right)\, dx  \, .
\end{eqnarray}

Similarly, the substrate potential energy $f_i$ and its expansion
take the form:

\begin{eqnarray}
f_i &=& \int_{L_0} \phi(d) (1+\dot{\xi}^2)^{1/2} \, dx - \phi(d_0)L\nonumber \\
&\sim& \int_{L_0} dx \, \left(\phi(d_0)+
\phi'(d_0)\xi+\frac{\phi^{(2)}(d_0)}{2}\xi^2 + \frac{\phi(d_0)}{2}
\dot{\xi}^2 \right.\nonumber \\
&& + \frac{\phi^{(3)}(d_0)}{6}\xi^3 + \frac{\phi'(d_0)}{2}\xi
\dot{\xi}^2 + \frac{\phi^{(4)}(d_0)}{24}\xi^4 +
\nonumber \\
&& \left.\frac{\phi^{(2)}(d_0)}{4}\xi^2\dot{\xi}^2 -
\frac{\phi(d_0)}{8}\dot{\xi}^4 + \cdots \right)- \phi(d_0)L\, .
\end{eqnarray}
Here $\phi'(d_0), \phi^{(2)}(d_0)\ldots$ denote the derivatives of
$\phi$ with respect to z-coordinate and are evaluated in the flat
state. Integration of the first order term in $\xi$ vanishes because
of volume conservation, equation (\ref{eqn:w:vol}). Furthermore, we
will choose the origin of x-coordinate such that integration ranges
from $-L_0/2$ to $L_0/2$. Different coordinate systems differ only
in small boundary terms, which are negligible if the system is large
enough. In the following discussion it will be clear in what sense
we mean by large enough. Minimizing $g'$ with respect to $L_0$, $L$
and surface undulation $\xi(x)$ and keeping only the lowest order
terms in $\xi$, we get the following equilibrium equations of state
\cite{mi-89}:

\begin{eqnarray}
\Pi_{ex}-\gamma_0 + \phi(d_0)+\theta &=& 0 \label{eqn:w:equi1} \\
\gamma_0 - \phi(d_0)+\theta + \frac{\partial(L
f_l(\sigma))}{\partial(L)}&=& 0
\label{eqn:w:equi2} \\
B \frac{d^4 \xi}{dx^4} - \left(\theta +\phi(d_0)\right) \ddot{\xi} +
\phi^{(2)}(d_0)\xi &=& 0   \, . \label{eqn:w:equi}
\end{eqnarray}
The Lagrange multiplier $\theta$ is related to the external pressure
$\Pi_{ex}$: $\theta = \gamma_0 -\Pi_{ex}-\phi(d_0)$. Moreover, the
equilibrium shape of the interface $\xi(x)$ must satisfy the above
differential equation. Using an {\it Ansatz} of sinusoidal
deformation: $\xi(x) = h \sin(qx)$, where $h$ is the amplitude and
$q$ is the wavenumber, we obtain:

\begin{equation}
B q^4 + (\gamma_0 - \Pi_{ex})q^2 + \phi^{(2)}(d_0)= 0  \, ,
\end{equation}
where $\theta=\gamma_0 - \Pi_{ex} - \phi(d_0)$ has been used (by
equation (\ref{eqn:w:equi1})). We have the following relation
between external pressure $\Pi_{ex}$ and wavenumber $q$:

\begin{equation}
\Pi_{ex} - \gamma_0 = B q^2 + \frac{\phi^{(2)}(d_0)}{q^2} \, .
\label{eqn:w:critical}
\end{equation}
An equivalent equation was obtained by Huang et al
\cite{Huang_2003}. Minimizing the right hand side of equation
(\ref{eqn:w:critical}) with respect to $q$, we get the smallest
external pressure $\Pi_c$ for buckling instability and critical
wavenumber $q_c$:

\begin{eqnarray}
q_c &=& \left(\frac{\phi^{(2)}(d_0)}{B}\right)^{1/4}  \nonumber \\
\lambda_c &=& 2\pi\left(B/\phi^{(2)}(d_0)\right)^{1/4} \nonumber \\
\Pi_{c} &=& \gamma_0 +2(B \phi^{(2)}(d_0))^{1/2}  \, .
\label{eqn:w:thresh}
\end{eqnarray}

If the gravitational energy of a liquid subphase (in air) is
considered, we may choose the flat state as reference and take
$\phi(d) = \rho g \xi^2/2$. The threshold external pressure $\Pi_c$
and critical wavenumber $q_c$ in this particular case are:

\begin{eqnarray}
\Pi_{c} &=& \gamma_0  + 2(B \rho g)^{1/2}
\nonumber \\
q_c &=& \left(\frac{\rho g}{B}\right)^{1/4} \label{eqn:w:mil}  \, .
\end{eqnarray}
The results in equation (\ref{eqn:w:mil}) were obtained by Milner et
al. in the paper \cite{mi-89}. Gravitational energy is important in
macroscopic scale with relatively thick liquid subphase. However, in
microscopic scale ($d <$100nm), different types of substrate
interaction between solid substrate and liquid subphase, e.g. van
der Waals interaction, become dominant, while gravity is negligible.
The buckling transition now leads to microscopic wrinkles. Our
result in equation (\ref{eqn:w:thresh}) is a generalization to this
microscopic range. The effect of different types of interaction will
be discussed in a later section.

\subsection{Second-order Buckling Transition}

In the previous section, we used the small deformation approximation
and expanded the Gibbs free energy to the lowest order in surface
displacement $\xi(x)$. A relation between external pressure
$\Pi_{ex}$ and wavenumber $q$ of wrinkles is obtained in equation
(\ref{eqn:w:critical}). In order to study the undulation amplitude
$h$ and possible order of the buckling transition, higher order
terms in the expansion should be included in the analysis. We are
particularly interested in finding out the existence conditions for
a continuous, second-order transition. From Landau's classical
theory of phase transition \cite{an-89}, a first-order
(discontinuous) transition will occur if the coefficient of the
fourth order term of free energy expansion with respect to the order
parameter ($\xi(x)$ in our case) is negative. In this section, we
will include terms up to the fourth order of $\xi$. The
inextensibility constraint of the surfactant layer yields:

\begin{equation}
L = \int_{L_0} \left(1 + \dot{\xi}^2 \right)^{1/2}\, dx  = const  \,
. \label{eqn:s:inex}
\end{equation}
Under the inextensibility constraint (\ref{eqn:s:inex}), we can drop
constant terms in expression (\ref{eqn:w:g'}) and minimize with
respect to the following functional of $L_0$ and $\xi(x)$:

\begin{equation}
g_1 = (\Pi_{ex}-\gamma_0)L_0 + f_b + f_i + \phi(d_0)L  \, .
\label{eqn:s:fun}
\end{equation}

The projected length $L_0$ in x-direction and the surface
deformation $\xi(x)$ are not independent variables. They are related
through the constraint (\ref{eqn:s:inex}). Assuming sinusoidal
deformation: $\xi(x)=h \sin(qx)$, it leads to an expression of $L_0$
in terms of $L$, $h$ and $q$. We will revisit the assumption of
incompressibility in the Discussion section. Inserting this
expression for $L_0$ into equation (\ref{eqn:s:fun}), we see that
the functional $g_1$ has the form of Landau free-energy expansion
\cite{an-89}. Firstly, we compute the expression for $L_0$. Expanded
to the fourth order in $\xi(x)$, i.e. $h \sin(qx)$, the constraint
(\ref{eqn:s:inex}) is:

\begin{eqnarray}
L &=& \int_{-L_0/2}^{L_0/2}dx (1 + \dot{\xi}^2)^{1/2} \nonumber \\
&\sim& \int_{-L_0/2}^{L_0/2}dx\left(1+ \frac{1}{2}\dot{\xi}^2 -
\frac{1}{8}\dot{\xi}^4 \right) \nonumber \\
&\sim& L_0\left(1 + \frac{1}{4}h^2 q^2 - \frac{3}{64}h^4 q^4\right)
\, . \label{eqn:s:noboundary}
\end{eqnarray}
In the last step, we keep only extensive terms, proportional to the
size of the system $L_0$, and neglect boundary terms. The boundary
terms are of the order of a wavelength of wrinkles: $2\pi/q$. The
approximation made in equation (\ref{eqn:s:noboundary}) is thus
essentially that the wavelength of wrinkles is much smaller than the
dimension of Langmuir system in the x-direction: $|q L|\gg 1$. If
this condition is satisfied, the approximation
(\ref{eqn:s:noboundary}) is valid and we get the following
expression for the projected length $L_0$:

\begin{eqnarray}
L_0 &=& L \left/ \left(1 + \frac{1}{4}\tilde h^2 -
\frac{3}{64}\tilde h^4\right) \right. \nonumber
\\
&\sim& L\left(1 - \frac{1}{4}\tilde h^2 + \frac{7}{64}\tilde
h^4\right)  \, , \label{eqn:s:L0}
\end{eqnarray}
where the slope amplitude $\tilde h = hq$ is a dimensionless
parameter and $|\tilde h| \ll 1$ in the small deformation
approximation. Similarly, expanding the functional $g_1$ in equation
(\ref{eqn:s:fun}) to the fourth order in $\xi$ and neglecting
boundary terms, we get:

\begin{eqnarray}
g_1 &\sim& L_0\left[-\theta + \frac{\tilde h^2}{4}\left(B q^2 +
\phi(d_0)+ \frac{\phi^{(2)}(d_0)}{q^2}\right) +
 \right. \nonumber\\
&&  \frac{\tilde h^4}{64}\left(-10B q^2 - 3\phi(d_0) + \frac{2
\phi^{(2)}(d_0)}{q^2} \right.\nonumber \\
&& \left.\left.+\frac{ \phi^{(4)}(d_0)}{q^4}\right) \right]\, ,
\end{eqnarray}
where $\theta =\gamma_0 - \Pi_{ex}-\phi(d_0)$ is the Lagrange
multiplier defined in the last section. $g_1$ depends on the
dimensionless variable $\tilde h$ and wavenumber $q$. Using the
expression of $L_0$ in equation (\ref{eqn:s:L0}):

\begin{eqnarray}
g_2 &\equiv& g_1/L \nonumber \\
&\sim& -\theta + \frac{\tilde h^2}{4}\left(B q^2 + \theta+\phi(d_0)+
\frac{\phi^{(2)}(d_0)}{q^2}\right)+ \nonumber \\
&&\frac{\tilde h^4}{64}\left(\frac{\phi^{(4)}(d_0)}{q^4} -
\frac{2\phi^{(2)}(d_0)}{q^2}-7\theta-7\phi(d_0) \right. \nonumber
\\
&& \left. - \, 14Bq^2\right) \, .
\end{eqnarray}

The equilibrium configuration minimizes the value of $g_2$. In the
above expression $g_2\rightarrow -\infty$ as $q\rightarrow \infty$.
It seems that the minimum value of $g_2$ doesn't exist. This paradox
is solved by noticing that the value of wavenumber $q$ can't vary
arbitrarily. In the small deformation approximation, we have assumed
that $\tilde h=hq$ is a small quantity. As a result, $q$ can't be
arbitrarily large. Furthermore, $q$ should be close to its critical
value $q_c$ near transition:

\begin{equation}
q = q_c\cdot (1+\delta) \, ,\label{eqn:s:q}
\end{equation}
where $\delta$ is a small dimensionless parameter and $\delta > 0$
if wavenumber $q$ is a continuous function of compression. Using the
expression (\ref{eqn:s:q}) for q and keeping the terms up to the
second order in $\delta$, we obtain:

\begin{equation}
g_2 \sim -\theta+ A_1 \tilde h^2 + A_2 \tilde h^4 \, ,
\end{equation}
where the coefficients $A_1$ and $A_2$ are quadratic functions of
$\delta$:

\begin{eqnarray}
A_1 &=& \frac{1}{4}(\Pi_{c}-\Pi_{ex}) +
\,\delta^2\pi \label{eqn:s:A1} \\
A_2 &=& \frac{1}{64}\left(a - 16 b - 7\theta-7\phi(d_0) \right) +
\frac{\delta}{16}\left(-a - 6b \right) + \nonumber \\
&& \frac{\delta^2}{32}\left(5a - 10b\right) \label{eqn:s:A2} \, ,
\end{eqnarray}
where the two parameters $b$ and $a$ are defined as:

\begin{eqnarray}
a &\equiv& \left(B \phi^{(4)}(d_0)\right)/\phi^{(2)}(d_0) \nonumber\\
b &\equiv& \left(B\phi^{(2)}(d_0)\right)^{1/2} \, .
\end{eqnarray}

The minimum value of $g_2$ exists for some positive $x$ and
$\delta$, only if the coefficients $A_1$ and $A_2$ satisfy the
inequalities:

\begin{equation}
A_1 < 0 \textrm{ ; } A_2>0 \, .\label{eqn:s:inequAs}
\end{equation}
Furthermore, if inequalities (\ref{eqn:s:inequAs}) are true, $g_2$
can achieve its minimum at $\tilde h_{min}$ and $\delta_{min}$:

\begin{eqnarray}
\tilde h_{min} &=& D_{\tilde h} \cdot \left(\Pi_{ex} - \Pi_c\right)^{1/2} \nonumber\\
\delta_{min} &=& D_{\delta} \cdot \left(\Pi_{ex} - \Pi_c\right) \, ,
\label{eqn:s:minxd}
\end{eqnarray}
where $D_{\tilde h}$ and $D_{\delta}$ are some positive
coefficients. Obviously, $\tilde h_{min}$ and $\delta_{min}$ are
positive when $\Pi_{ex}>\Pi_{c}$ and approach zero, as the external
pressure $\Pi_{ex}$ approaches its threshold value $\Pi_c$ from
above. In other words, the values of $\tilde h_{min}$ and
$\delta_{min}$ can be made arbitrarily small by approaching
transition point. As a result, the terms involving $\delta$ in $A_1$
and $A_2$ are of higher order and can be neglected in comparison
with the finite constant term. The inequalities
(\ref{eqn:s:inequAs}) are reduced to:

\begin{eqnarray}
\Pi_{c}-\Pi_{ex} & < & 0 \\
a - 16 b - 7\theta-7\phi(d_0) &>& 0 \, .
\end{eqnarray}
The first inequality will be true if $\Pi_{ex} > \Pi_{c}$. Using the
expression of $\gamma$, $a$ and $b$, we can rewrite the second
inequality as:

\begin{equation}
\left(B \phi^{(4)}(d_0)\right)/\phi^{(2)}(d_0) -
2\left(B\phi^{(2)}(d_0)\right)^{1/2} > 7(\Pi_c - \Pi_{ex}) \, .
\end{equation}
Since the right hand side approaches zero from below as $\Pi_{ex}$
decreases to $\Pi_c$, we get the condition:

\begin{equation}
\phi^{(4)}(d_0)> 2 \left[\frac{\phi^{(2)}(d_0)^{3}}{B}\right]^{1/2}
\, , \label{eqn:s:inequphi}
\end{equation}
where we have used the expression (\ref{eqn:w:thresh}) for $q_c$. In
order to have a second-order buckling transition, the inequality
(\ref{eqn:s:inequphi}) is the condition that must be satisfied by
the substrate potential $\phi(d)$. The above criterion for the order
of buckling transition doesn't change after including the second
harmonic term $\alpha h \sin(2qx)$. Because $\alpha$ doesn't affect
$A_1$ and enters $A_2$ as $\alpha^2 \tilde h^2$, the second harmonic
term affects $g_2$ only at order $\tilde h ^6$ or higher. As a
result, it does not change the criterion in inequality
(\ref{eqn:s:inequphi}). In the case of gravitational potential
energy of liquid subphase: $\phi(d)=\rho g \xi^2 /2$, we have
$\phi^{(2)}(d_0)=\rho g>0$ and $\phi^{(4)}(d_0) = 0$. The relation
(\ref{eqn:s:inequphi}) cannot be satisfied. Therefore, we will have
a first-order buckling transition in this case \cite{mi-89}.
However, the relation (\ref{eqn:s:inequphi}) may be satisfied for
some types of substrate potentials. We will discuss its explicit
forms in the next section and consider the possibility of a
second-order buckling transition for different types of interaction.
Although our conditions are necessary for a continuous transition,
they are not completely sufficient. To show that no discontinuous
transition occurs, we would have to show that {\it no} displacement
$\xi(x)$ has a $\Pi_{min} < \Pi_c$. We considered only small and
harmonic $\xi(x)$. \\

\section{EXAMPLE POTENTIALS}

In this section, we will consider some examples of substrate
potentials and find out the corresponding existence conditions for
second-order buckling transition.

\subsection{Non-retarded Van der Waals Interaction}

Non-retarded van der Waals interaction between liquid subphase and
solid substrate takes the form \cite{de-03}:

\begin{equation}
\phi(d) = \frac{A}{12\pi d^2} \, .
\end{equation}
Here $A$ is the Hamaker constant and has the dimension of energy. It
can be positive or negative depending on the properties of liquid
subphase and solid substrate. If $A$ is positive, van der Waals
interaction leads to an effective repulsion between liquid-air and
liquid-substrate interfaces and favors a thicker liquid film, i.e.
larger $d$. On the other hand, if $A < 0$, the liquid film can be
unstable. Spontaneous fluctuations may rupture the liquid film via
spinodal dewetting \cite{vi-67}. The value of $A$ is typically in
the range of $10^{-20}$ J to $10^{-19}$ J \cite{de-03}. From
equation (\ref{eqn:w:thresh}), we have the critical wavenumber $q_c$
and wavelength $\lambda_c$:

\begin{eqnarray}
q_c &=& \frac{1}{d_0}\left(\frac{A}{2\pi B}\right)^{1/4} \nonumber
\\
\lambda_c &=& 2 \pi d_0 \left(\frac{2\pi B}{A}\right)^{1/4} \, .
\label{eqn:d:vanlc}
\end{eqnarray}
Thus $q_c$ and $\lambda_c$ exist only in the case $A > 0$, i.e. for
stable liquid subphase. The second-order buckling transition
requirement (\ref{eqn:s:inequphi}) reduces to:

\begin{equation}
A < 200 \, \pi B \label{eqn:d:van} \, ,
\end{equation}
which doesn't depend on the thickness of the liquid subphase $d_0$.
This independence of thickness can be understood by noticing that as
we increase the thickness of liquid subphase $d_0$, van der Waals
potential energy density decreases as $A/d_0^2$. Meanwhile, the
local curvature of surfactant layer $C$ is of the order of $1/d_0$,
so bending free energy density varies as $B/d_0^2$, which has the
same functional form as van der Waals interaction. As a result, we
expect the criterion doesn't depend on the thickness of the liquid
subphase $d_0$ and is a relation between the Hamaker constant $A$
and bending modulus $B$. As long as $A$ and $B$ satisfy condition
(\ref{eqn:d:van}), microscopic wrinkles can be formed through a
second-order buckling transition. For a lipid monolayer $B \gtrsim
10 kT$ \cite{da-91}, which is $4 \times 10^{-20}$ J at 25
centigrade. The smallest $\lambda_c$ compatible with equation
(\ref{eqn:d:van}) is:

\begin{equation}
\lambda_{c\,min} = 2 \pi d_0/\sqrt{10} \, \thickapprox \, 2.0 \, d_0
\, .
\end{equation}

The wavelength $\lambda_c$ increases only gradually from the minimum
because of the weak dependence on $B/A$ in equation
(\ref{eqn:d:vanlc}). For practical purposes, the buckling wavelength
is confined to scales of order $d_0$. In the microscopic range that
we are discussing, $d_0 < 100nm$. Typically, the dimension of the
Langmuir system $L$ in horizontal direction is about 1mm, so the
condition of a large enough system $|q_c L |\gg 1$ is satisfied. \\

\subsection{Charged Double-layer Interaction}

The charged double-layer interaction has the form \cite{is-85}:

\begin{equation}
\phi(d) = \phi_0 \exp{(-\kappa d)} \, ,
\end{equation}
where $1/\kappa$ is Debye screening length. The coefficient $\phi_0$
is a constant depending on zeta-potentials of two surfaces and
electrolyte concentration in between. For a 1:1 electrolyte,
$\phi_0$ can be written as:

\begin{equation}
\phi_0 = \frac{64kT \gamma_1\gamma_2 C_s}{\kappa} \, .
\label{eqn:d:weak}
\end{equation}
In equation (\ref{eqn:d:weak}), T is the temperature and $C_s$ is
the concentration of electrolyte in the bulk. Moreover $\gamma_i$
(i=1,2) is related to zeta-potentials $ \zeta_{0i} $ of the two
surfaces \cite{is-85}:

\begin{equation}
\gamma_i = \textrm{tanh}(e \zeta_{0i} /4kT) \, .\label{eqn:d:gammai}
\end{equation}

The potential $\phi_0$ is positive if two electrical surfaces have
charges of the same sign and repel each other. In the case $\phi_0$
is negative, two electrical surfaces attract each other; this may
rupture the liquid film via unstable modes of undulation. The above
equation is valid in the weak overlap approximation , in which the
overlap between electrical double layers is small \cite{da-93}:

\begin{equation}
\kappa \cdot d >1  .
\end{equation}
From equation(\ref{eqn:w:thresh}), we have the critical wavenumber
$q_c$ and wavelength $\lambda_c$:
\begin{eqnarray}
q_c &=& \left(\frac{\kappa^2\phi_0}{B}\exp{(-\kappa
d_0)}\right)^{1/4} \nonumber
\\
\lambda_c &=& 2 \pi \left(\frac{B}{\kappa^2\phi_0}\exp{(\kappa
d_0)}\right)^{1/4} \, .
\end{eqnarray}
Thus $\lambda_c$ and $q_c$ exist only in the case $\phi_0
>0$, i.e. for stable liquid subphase. The second-order buckling transition condition
(\ref{eqn:s:inequphi}) takes the form:

\begin{equation}
d_0 > \frac{1}{\kappa}\ln{\left(\frac{4\phi_0}{\kappa^2B }\right)}
\, . \label{eqn:d:double}
\end{equation}

The smallest $\lambda_c$ compatible with condition
(\ref{eqn:d:double}) is:

\begin{equation}
\lambda_{c \, min} = 2\sqrt{2} \pi \kappa^{-1} \, .
\end{equation}
It is on the order of the Debye screening length $\kappa^{-1}$. The
wavelength $\lambda_c$ is an increasing function of $d_0$. For
example, we consider a negatively-charged mica substrate and a
negatively-charged conventional liposome (lecithin/cholestrol 6:4
molar ratio). As an electrolyte, we consider NaCl at concentration
$C_s$ = 0.001 mol/L. In this case, the Debye screening length is
$\kappa^{-1} \approx 10$ nm \cite{is-85}. At pH 5.8, the zeta
potentials of mica surface \cite{jo-81} and conventional liposomes
are \cite{la-93}:

\begin{equation}
\zeta_{0 \,mic} = -104 \textrm{mV} \textrm{\, and \,} \zeta_{0 \,
lip} = -20 \textrm{mV}  \, .
\end{equation}

The value of $\phi_0$ is calculated as:

\begin{equation}
\phi_0 \thickapprox 2.2 \times 10^{-4} \, \textrm{(J/m$^2$)}  \, .
\end{equation}
By the estimation of the bending modulus $B \gtrsim 10 kT$
\cite{da-91}, the condition (\ref{eqn:d:double}) reduces to:

\begin{equation}
d_0 > 8.1 \textrm{nm}  \, .
\end{equation}

\subsection{Sharma Potential}

The Sharma potential is used widely in studies of wetting phenomena
between different liquid thin films and solid substrates
\cite{sha-93, lee-03, oss-88}, e.g. water on mica. It includes both
the apolar(Lifshitz-van der Waals) and polar interactions and has
the form:

\begin{equation}
\phi(d) = S^{AP}\cdot\left( \frac{d_c^2}{d^2}\right)+S^{P}\cdot
\exp{\left[\frac{d_c-d}{l}\right]} \, ,
\end{equation}
where $S^{AP}$ and $S^{P}$ are apolar and polar contribution to the
spreading coefficient. $l$ is the correlation length for polar
liquid and $d_c$ is the Born repulsion cutoff length \cite{sha-93}.
In this case, the critical wavenumber $q_c$ and wavelength
$\lambda_c$ have the form:

\begin{eqnarray}
q_c &=& \frac{1}{d_0 B^{1/4}}\left(6 S^{AP} d_c^2+\frac{S^P
d_0^4}{l^2}
\exp{[(d_c-d_0)/l]} \right)^{1/4} \nonumber\\
\lambda_c &=& 2 \pi d_0 B^{1/4}\left(6 S^{AP} d_c^2+\frac{S^P
d_0^4}{l^2} \cdot \exp{[(d_c-d_0)/l]} \right)^{-1/4}\textrm{.}
\end{eqnarray}

The existence requirement of second-order buckling transition
yields:

\begin{eqnarray}
120  S^{AP}d_c^2 l^4 &>&- S^p d_0^6
\exp{\left(\frac{d_c-d_0}{l}\right)}+ 2\,l/\sqrt{B}\left[6 S^{AP} d_c^2 l^2 \right.\nonumber \\
&& \left. + \, d_0^4 S^p \exp{\left(\frac{d_c-d_0}{l}\right)}
\right]^{3/2}    \, . \label{eqn:d:sharma}
\end{eqnarray}

For example, we consider a lipid monolayer, with bending modulus $B
\gtrsim 10 kT$, sitting on top of water with a mica substrate below.
The numerical values of coefficients are $S^{AP} = 20mN/m$, $S^{P} =
48mN/m$, $l=0.6 nm$ and $d_c=0.158nm$ \cite{lee-03}. The wavelength
$\lambda_c$ is an increasing function of $d_0$. In the microscopic
range, where the thickness of water subphase $d_0 < 100nm$, we get
an upper bound for $\lambda_c$:

\begin{equation}
\lambda_c \lesssim 1.2 \, \mu m  \, .
\end{equation}
Obviously, the condition $|q_c L|\gg 1$ is guaranteed. Furthermore,
the inequality (\ref{eqn:d:sharma}) is true for any positive value
of $d_0$. In other words, the buckling transition will always be
second-order if the sharma potential correctly describes the
interaction between water subphase and a mica substrate. The
following dimensionless quantity changes very slowly with the value
of $d_0$:

\begin{equation}
\eta \equiv B^{1/4}\left(6 S^{AP} d_c^2+\frac{S^P d_0^4}{l^2} \cdot
\exp{[(d_c-d_0)/l]} \right)^{-1/4}   \, .
\end{equation}

In the range of $d_0$ between $1nm$ and $100nm$, $\eta \in$ [0.8,
1.9]. Thus the critical wavelength $\lambda_c$ can be written as:

\begin{eqnarray}
\lambda_c &=& 2 \pi d_0 \eta \nonumber \\
&\gtrsim& 5.0 \, d_0  \, .
\end{eqnarray}
For the estimation in the last step, $\eta$ takes the smallest value
0.8 in this range. \\

\section{NUMERICAL RESULTS}

\begin{figure}[!htp]
\begin{center}
\includegraphics[angle=-90, width=3.7in]{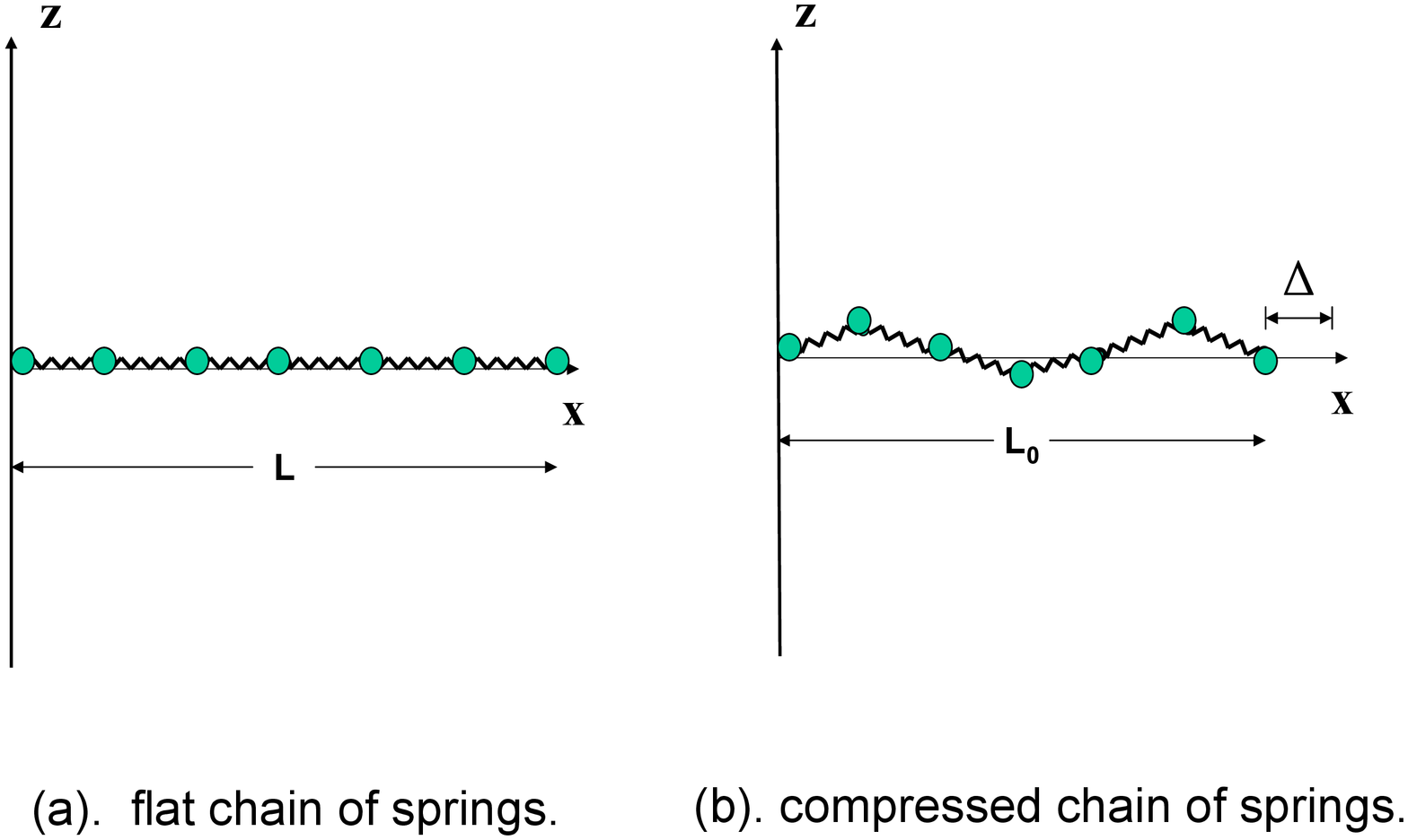}
\caption{Discrete model of one dimensional chain of springs used in
the simulation. (a). Flat state with no compression. (b). Buckled
state with large enough compression.} \label{fig:3:springs}
\end{center}
\end{figure}

In order to verify our results and explore the predicted wrinkling
phenomena concretely, we have done a discrete numerical
implementation of the system. The numerical simulation was carried
out using the Mathematica program. We modeled the surfactant layer
by a one-dimensional chain of nodes connected by springs with
un-stretched length $a$ and spring constant $k$. The un-stretched
length $a$ is set to be 1 in the simulation for convenience. In
order to impose the inextensibility constraint, we set the spring
constant $k$ to be a very large value. A bending energy of
$B\theta_{i,(i+1)}^2/2$ is assigned to every pair of adjacent
springs, where $\theta_{i,(i+1)}$ is the angle between these two
springs along the chain direction. The total bending energy $f_b$ is
a sum of all pairs along the chain. In the numerical simulation we
set the free liquid-air surface tension in our theory $\gamma_0 =
0$. The substrate potential is discretized correspondingly by
replacing the integral with a summation along the chain. With no
compression the chain adopts a flat configuration lying on the x
axis. In the simulation, the first node is fixed at the origin. In
order to reduce the influence of boundary effects in a finite system
used in the simulation, we fix the z-coordinate of the last node to
be zero, while its x-coordinate was determined by the amount of
compression $\Delta$, as shown in Fig.\ref{fig:3:springs}. All the
other nodes are movable both in x and z directions in the process of
minimization. The total free energy $g^*$ of this discrete model
takes the form:

\begin{figure}[!htp]
\begin{center}
\includegraphics[angle=-90, width=3.6in]{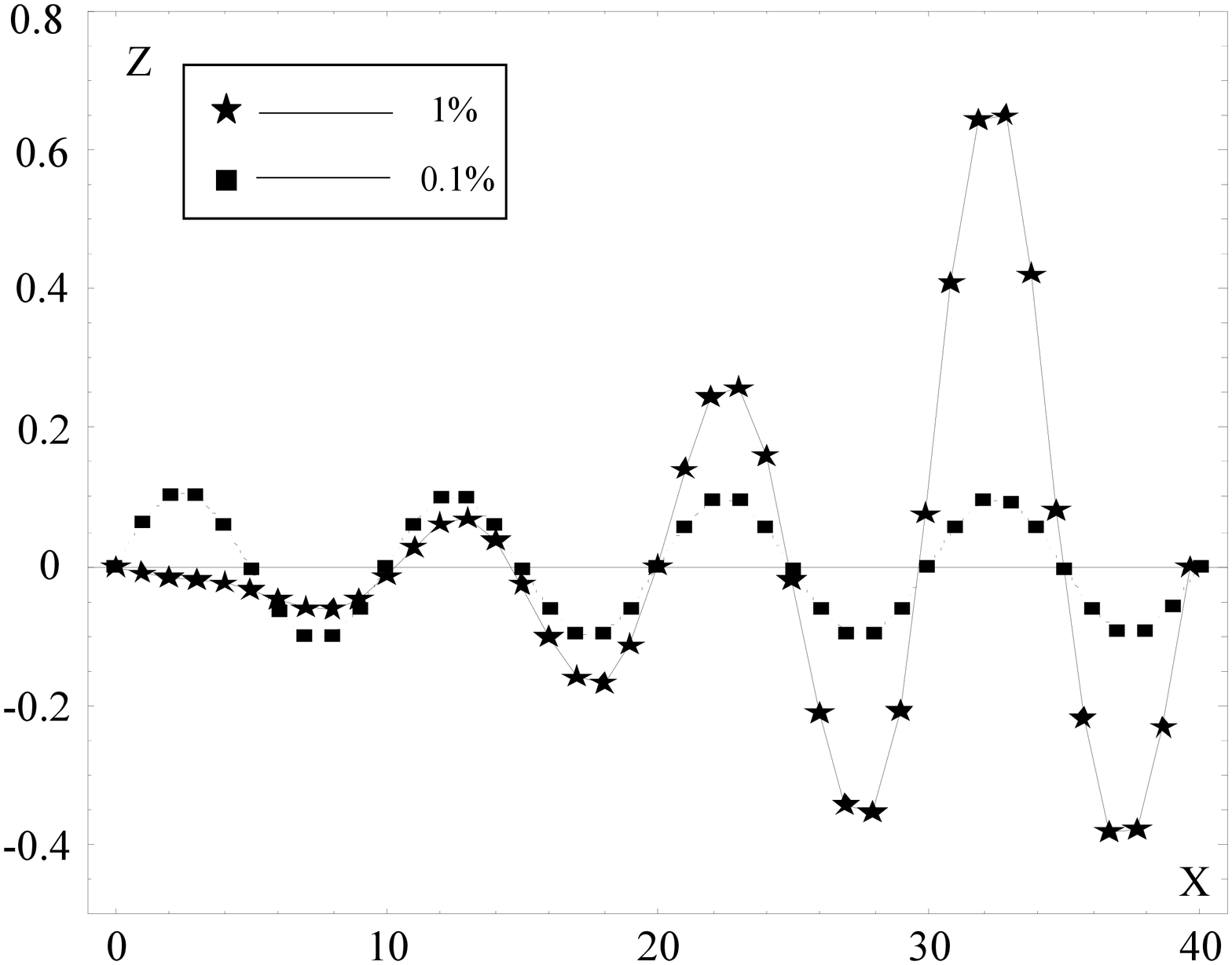}
\caption{Configurations with 41 nodes as compression $\Delta$
changes. The box in the graph shows the amount of compression. Van
der Waals interaction is used in this simulation. Parameter values:
$A=200$, $B=1$, $d_0=3.78036$, $k=10^{11}$. The predicted wavelength
of wrinkling: $\lambda_c = 10$. The nonuniformity of the amplitude
with 1\% compression is due to nonlinear effects as approaching a
possible wrinkle-to-fold transition \cite{luka}. The position of a
fold could change for different initial conditions of minimization.}
\label{fig:3:config1}
\end{center}
\end{figure}

\begin{eqnarray}
g^* &=& \Pi_{ex}L_0 + f_b + f_i + f_k \nonumber \\
    &=& \Pi_{ex}L_0 + g_1^* \, , \label{eqn:3:gibbs}
\end{eqnarray}
where $f_b$, $f_i$ and $f_k$ are bending energy, substrate potential
energy and elastic energy stored in the springs. The quantity
$g_1^*$  is the sum of these. For the flat reference state, the
above three energy terms are zero. So the total free energy for the
flat state is:

\begin{equation}
g^* = \Pi_{ex}L \, ,
\end{equation}
where $L$ is the total length of the chain. The compression is
$\Delta = L - L_0$. At each fixed amount of compression $\Delta$, we
minimized the value of $g_1^*(\Delta)$ and computed the smallest
external pressure $\Pi_{min}(\Delta)$ needed to reach this
compression via the following equation:

\begin{equation}
\Pi_{min}(\Delta) = \frac{g_{1min}^*(\Delta)}{\Delta}  \, .
\end{equation}

\begin{figure}[!htp]
\begin{center}
\includegraphics[angle=-90, width=3.6in]{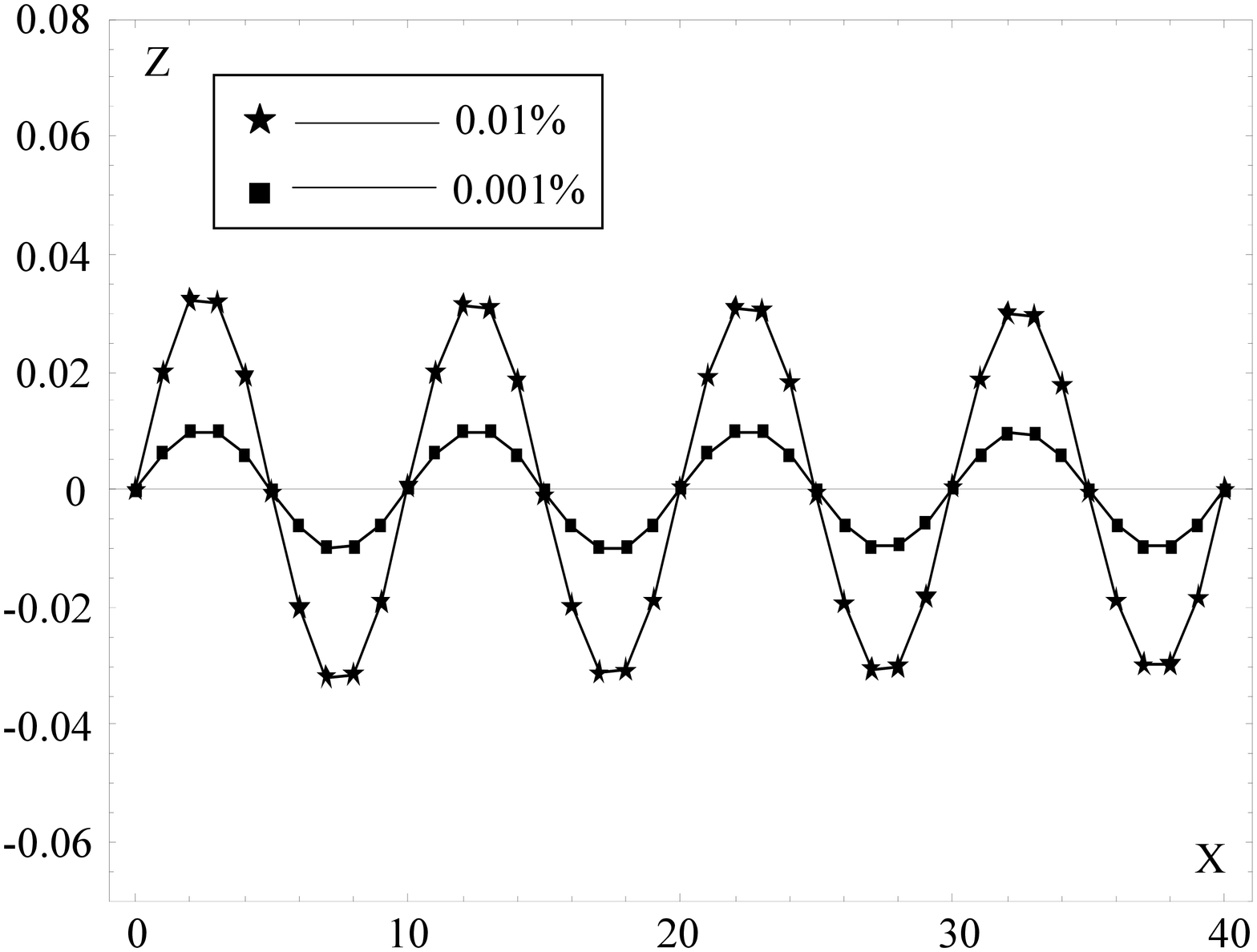}
\caption{Configurations with 41 nodes as compression $\Delta$
changes. The box in the graph shows the amount of compression.
Parameters values are the same as in Fig.3.} \label{fig:3:config2}
\end{center}
\end{figure}

We chose a value of $d_0$ such that the predicted $\lambda_c$ was
commensurate with the system, namely $\lambda_c =10$. Then starting
from a finite amount of compression, e.g. $\Delta=1\%$, we gradually
lowered the value of $\Delta$. If our theory is correct, the chain
will approach a sinusoidal shape with wavelength $\lambda_c$.
Moreover $\Pi_{min}(\Delta)$ should approach $\Pi_c$ as $\Delta$
goes to zero. The order of buckling transition can be deduced from
the functional shape of $\Pi_{min}(\Delta)$. For a second-order or
continuous transition, $\Pi_{min}(\Delta)$ is a monotonic increasing
function of $\Delta$. There is no jump when crossing the transition
point. In the case of a first-order transition, $\Pi_{min}(\Delta)$
is not monotonic. It has a minimal value $\Pi_{min}^*$ at a nonzero
compression $\Delta^*$. As a result, as soon as $\Pi_{ex}$ exceeds
$\Pi_{min}^*$, the configuration will jump to this finite amount of
compression showing the property of a first-order buckling.

\begin{figure}[!htp]
\begin{center}
\includegraphics[angle=-90, width=3.6in]{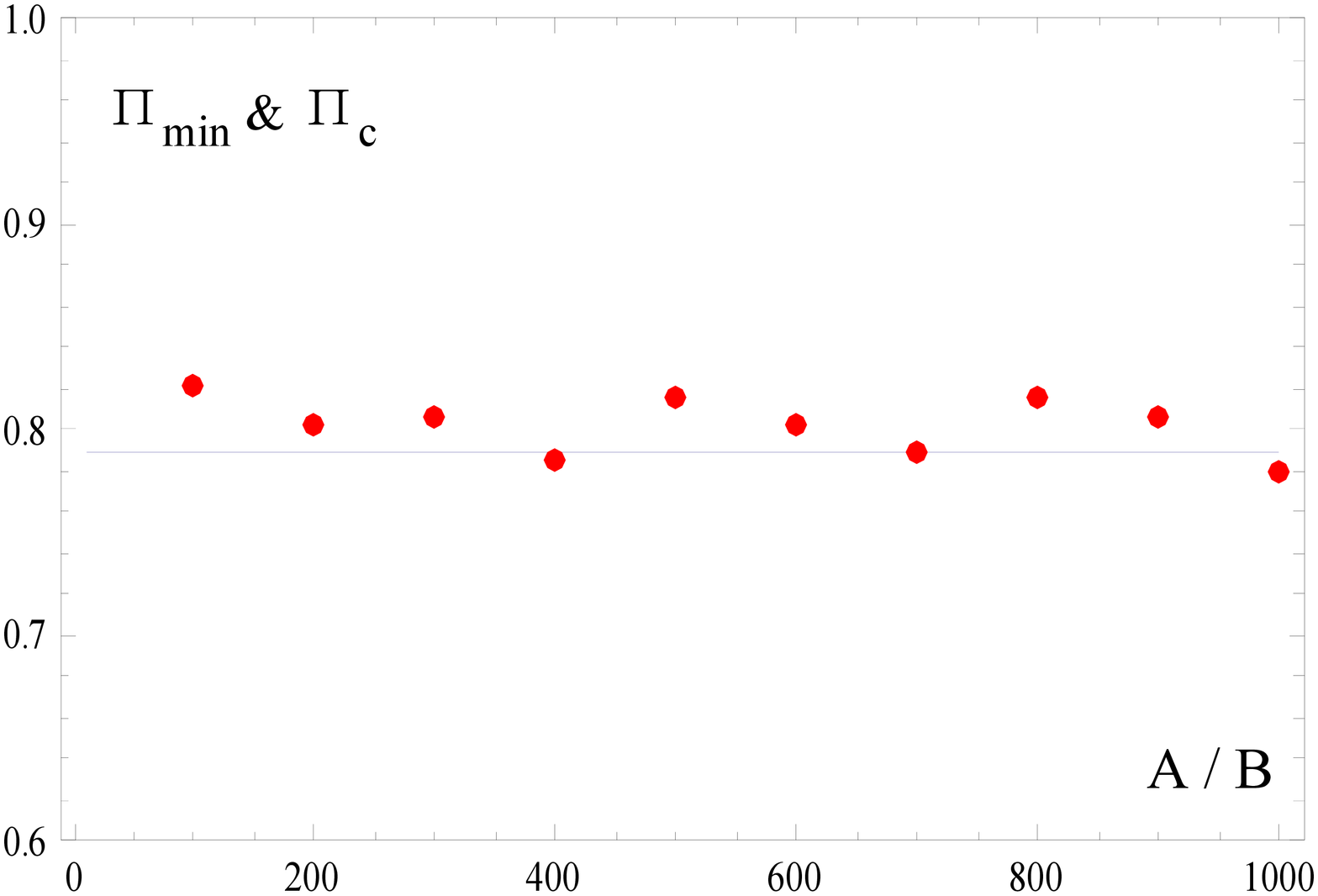}
\caption{Theoretically predicted $\Pi_c$ and $\Pi_{min}$ from
simulation as functions of ratio $A/B$. $B$ is fixed at $1$. Solid
line represents prediction from theory. Discrete points are values
of $\Pi_{min}$ from the simulation.} \label{fig:3:Disvstheo}
\end{center}
\end{figure}

A typical sequence of chain configurations with 41 nodes as we
changed the amount of compression is shown in
Fig.\ref{fig:3:config1} and \ref{fig:3:config2}.
Fig.\ref{fig:3:Disvstheo} shows the agreement between predicted
values of $\Pi_c$ and $\Pi_{min}$ from simulation for several cases.
It is noticed that $\Pi_c$ is a constant function of $A$, if the
buckling wavelength $\lambda_c$ is fixed.

Verifying the predicted transition from continuous to discontinuous
wrinkling proved to be more subtle. Even though the ratio $A/B$ is
nearly a factor of 2 above the predicted threshold, we didn't see
any evidence of discontinuous buckling using discrete model. To
understand this required a second numerical method. As we discuss
below, it reveals that the discontinuity is too weak to have been
seen in the discrete model.

\begin{figure}[!htp]
\begin{center}
\includegraphics[angle=-90, width=3.6in]{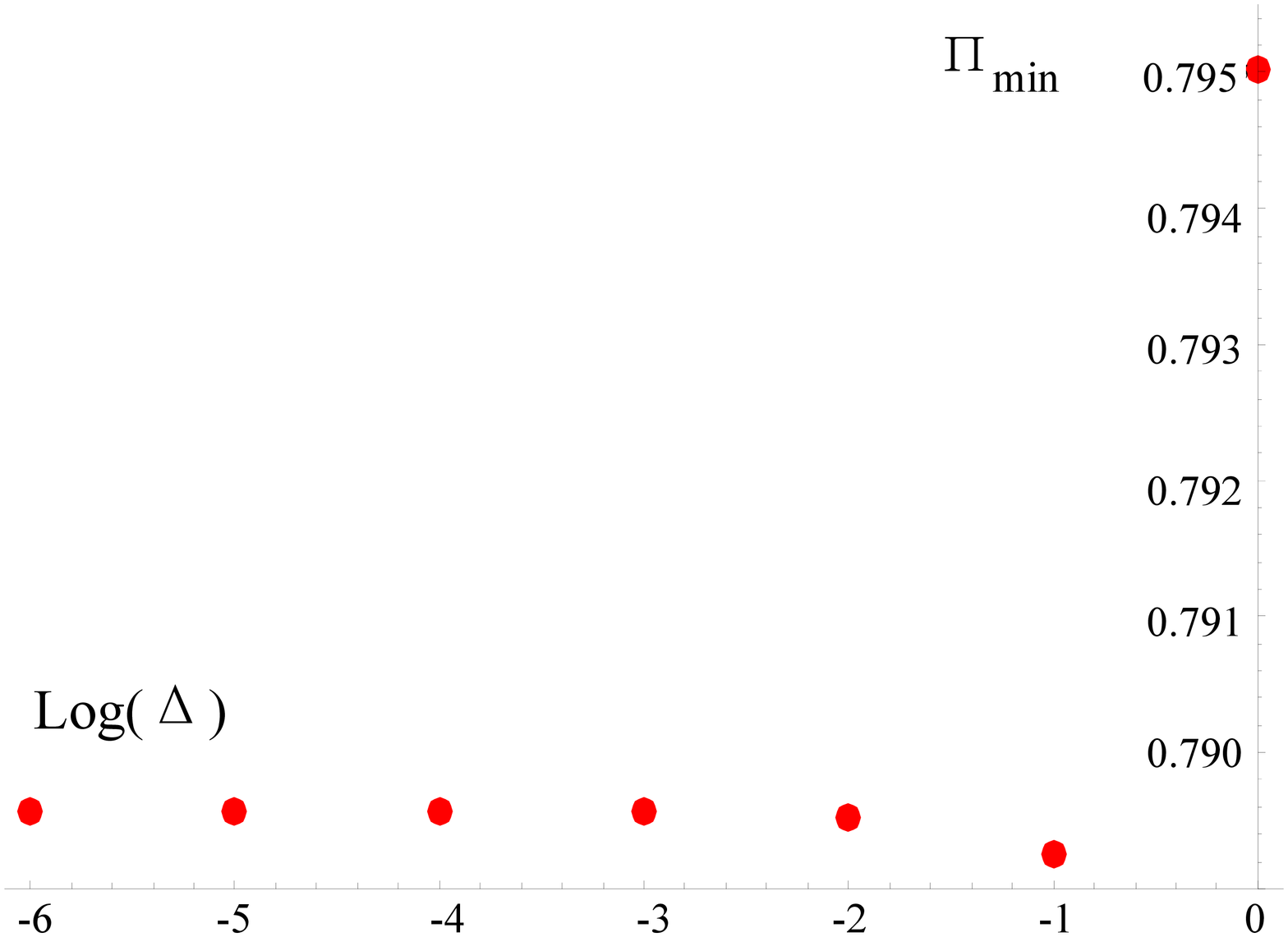}
\caption{$\Pi_{min}$ versus Log$(\Delta)$ for very weak first-order
transition in the case of van der Waals interaction. $A=1000$,
$B=1$, $d_0=5.65295$. The critical wavelength is $\lambda_c =10.$}
\label{fig:3:continous}
\end{center}
\end{figure}

The second numerical method is based on a continuous model. To
simplify the calculation, we took a single sinusoidal mode as {\it
Ansatz}: $h \sin{(qx)}$. For convenience, only one period of
wrinkling is included in the continuous model, while a real system
is composed of many copies of it. The wavelength $\lambda$ changes
as we change the amount of compression $\Delta$: $\lambda =
\lambda_c - \Delta$. Using a similar approach as the discrete model
described above, we can compute $\Pi_{min}(\Delta)$ for each amount
of compression. The order of transition is still determined by the
functional shape of $\Pi_{min}(\Delta)$. This time not only the
values of $\Pi_c$, but also the order of buckling are in good
agreement with our theoretical prediction. The paradox with discrete
model is also explained. As it is shown in
Fig.\ref{fig:3:continous}, the first-order transition is very weak
in the case of van der Waals interaction. With $A/B=1000$,
$\Pi_{min}$ has a minimal value at $1\%$ compression with a $0.6\%$
change in $\Pi_{min}$. Thus the buckling transition is first-order.
However, such a small change in $\Pi_{min}$ cannot be detected in
the above discrete model. Based on the above numerical results, our
theory
makes good prediction for the substrate-induced buckling transition.\\

\section{DISCUSSION}

In the previous sections, a substrate-bending model was constructed.
Here we discuss the implications of this model. Using our model, we
made prediction about wrinkling wavelength $\lambda_c$ with large
enough compression $\Pi_{ex} > \Pi_c$ and the order of buckling
transition. In order for our mechanism to account for wavelengths of
hundreds of nanometers, the trapped fluid layer itself would have to
be many nanometers thick. During the transfer of a monolayer to
substrate, such thick fluid layers usually exist. The compressive
stress required to buckle the surfactant layer could be developed
during this transfer and rapid drying after deposition.

The wrinkling mechanism predicted here is expected in any supported
monolayer or bilayer systems with sufficient compression. A
well-defined wrinkling wavelength $\lambda_c$ is given in terms of
the thickness of the subphase $d_0$, the bending modulus of the
surfactant layer $B$ and functional form of the substrate potential
$\phi(d)$. Information about these microscopic variables is embedded
in the experimentally observed wavelength. It is especially useful
if one can control the external pressure $\Pi_{ex}$. The properties
of gravitational wrinkles have been experimentally studied
\cite{luka}. Gravitational buckling appears to give rise to a
strongly first-order wrinkling-to-folding transition, which
contrasts with our very weak first-order transition in the case of
van der Waals interaction. Some other forms of substrate potential
$\phi(d)$ may give rise to a stronger first-order transition.

These surfactant layers are potentially subject to another kind of
instability different from the extensive wrinkling investigated
here. The boundary conditions may be such that the boundary region
buckles while the bulk of the layer is still in a stable state.
These boundary induced deformations of surfactant layers have been
studied \cite{Safran_book,Safran_2002}. Such boundary buckling was
an important factor in our discrete simulation. It prevented us from
studying arbitrary wavelengths. Also our methods only allowed us to
study the region of incipient instability. There may be interesting
phenomena analogous to the gravitational wrinkling-to-folding
transition that we have missed.

Inextensibility of the surfactant layer is an important assumption
in our theory. If this constraint were released, the system would
have a compression mode as an extra degree of freedom to store
elastic energy besides bending mode studied above. As shown in the
appendix, in the limit of small deformation approximation, finite
compressibility influences only fourth order and higher terms in the
free energy, so it doesn't change the expressions of the threshold
external pressure $\Pi_c$ and the critical wavelength $\lambda_c$.
Moreover, if the system were not too far away from the transition
threshold between first-order buckling and second-order buckling,
the inextensibility would be a good approximation in the experiments
of interest here.

\section{CONCLUSION}

\noindent As supported monolayers and bilayers become more commonly
studied, we expect that the type of wrinkling predicted here will be
observed and used to infer local properties, such as substrate
depth, bending modulus of surfactant layer and etc. It will be of
interest to see how such buckling occurs in time, and what
counterparts of the wrinkling-to-folding transition might exist.

\begin{acknowledgments}
We would like to thank Prof. Ka Yee Lee and her group members,
Guohui Wu and Shelli Frey, for providing experimental data. We also
want to thank Luka Pocivavsek and Enrique Cerda for insightful
discussions. This work was supported in part by the National Science
Foundation's MRSEC program under Award Number DMR-0213745 and by the
US-Israel Binational Science Foundation.
\end{acknowledgments}

\appendix

\section{}

To estimate the influence of compressibility, we assume that $t$ is
the thickness of the surfactant layer. The bending modulus $B$
varies as $t^3$, while the compressibility modulus $K$ is
proportional to $t$ \cite{Seifer_1994}: $B/K \simeq t^2$. The fourth
order term of bending free energy takes the form:

\begin{equation}
b_4 = \frac{5B}{4}\int_{L_0}  \dot{\xi}^2 \, \ddot{\xi}^2 dx \, .
\end{equation}

In the limit of small deformation approximation, derivatives of
$\xi$ can be approximated as:

\begin{eqnarray}
\dot{\xi} &\sim& h/\lambda_c \nonumber \\
\ddot{\xi} &\sim& h/\lambda_c^2 \label{eqn:app:small} \, .
\end{eqnarray}
Inserting equation (\ref{eqn:app:small}) into the expression of
$b_4$, we get:

\begin{equation}
b_4 \sim \frac{5 \,B \,h^4 \,L_0}{4\, \lambda_c^6} \, .
\label{eqn:app:b4}
\end{equation}

The compression free energy takes the form:

\begin{equation}
E_k = \frac{K}{2} \int_{L_0}\eta^2 \, \left(1 +
\dot{\xi}^2\right)^{1/2} \, dx \, ,
\end{equation}
where $K$ is the compressibility modulus. $\eta$ is the percentage
change of the length of surfactant layer. Choosing the configuration
of the surfactant layer just before buckling as a reference state,
we have:

\begin{eqnarray}
\eta &=& 1 - \frac{\left(\int_{L_0} (1 + \dot{\xi}^2)^{1/2} \, dx
\right)} {L_c} \nonumber \\
&\simeq& \frac{\Pi_{ex}-\Pi_{c}}{K} \label{eqn:app:eta} \, ,
\end{eqnarray}
where $L_c$ is the length of the surfactant layer just before
buckling transition. If buckling is not allowed, the compressive
strain $\eta$ is evidently given by the fractional decrease in $L$,
viz $(L_c-L_0)/L_c$. However, if buckling occurs, this strain can
only decrease. As a result, the following relation holds:

\begin{equation}
\eta  <  \frac{L_c - L_0}{L_c} \sim \tilde h^2 \, ,
\end{equation}
where $\tilde h$ is the slope amplitude. Thus, the expansion of
compression free energy in terms of $\tilde h$, being quadratic in
$\eta$, has only a fourth order term or higher. As a result, it
doesn't affect the expressions of $\Pi_c$ and $\lambda_c$. To the
lowest order approximation, $E_k$ can be written as:

\begin{equation}
E_k \sim c_4 = \frac{K \, \eta^2 \, L_0}{2} \, .
\end{equation}

Comparing $b_4$ and $c_4$, we get a criterion of inextensibility:

\begin{equation}
\eta \ll \frac{t}{\lambda_c} \cdot \tilde h^2  \label{eqn:app:inex}
\, ,
\end{equation}
where we have used the relation $B/K \simeq t^2$. By equation
(\ref{eqn:app:eta}), we have:

\begin{eqnarray}
\eta &\simeq& \frac{\Pi_{ex}-\Pi_c}{K} \nonumber \\
&=& \frac{\Pi_{ex}-\Pi_{c}}{A_2} \cdot
\frac{A_2}{K} \nonumber \\
&\sim& \tilde h^2 \cdot A_2/K \, ,
\end{eqnarray}
where the expression of $A_2$ was given in equation
(\ref{eqn:s:A2}). Inserting into the above criterion of
inextensibility, we have:

\begin{equation}
\frac{A_2}{K} \ll \frac{t}{\lambda_c} \label{eqn:app:inexA2} \, .
\end{equation}

As $\Pi_{ex}$ approaches $\Pi_c$ from above, $A_2$ can be
approximated as:

\begin{eqnarray}
A_2 &\simeq& \frac{1}{64}\left(B\, \phi^{(4)}(d_0)/\phi^{(2)}(d_0) -
2 (B\phi^{(2)}(d_0))^{1/2}\right) \nonumber \\
&=& \frac{1}{64}\left(B \, \phi^{(4)}(d_0)/\phi^{(2)}(d_0) - 8\pi^2
\, B /\lambda_c^2\right) \label{eqn:app:A2} \, ,
\end{eqnarray}
where we have used the expression of $\lambda_c$ in the last step.
We require $A_2 > 0$ in order for the transition to be second-order.
Thus, the first term in equation (\ref{eqn:app:A2}) must dominate
the second. However, if the system were not too far away from the
transition threshold between first-order buckling and second-order
buckling, two terms in equation (\ref{eqn:app:A2}) would have
comparable order of magnitude. In such a case, we can simplify the
criterion (\ref{eqn:app:inexA2}) as:

\begin{eqnarray}
\frac{A_2}{K} \sim \frac{B}{\lambda_c^2\, K} &\ll&
\frac{t}{\lambda_c} \nonumber \\
\textrm{i.e. } \, \, \frac{t}{\lambda_c} &\ll& 1 \, .
\end{eqnarray}
The above criterion is always satisfied in the experimental systems
that we are interested in. For example, suppose the monolayer
thickness $t$ is about $2$nm and the wrinkling wavelength
$\lambda_c$ is more than $100$nm. In such a case the approximation
of inextensibility is valid. In cases of very anharmonic potentials
where $\phi^{(4)}(d_0) \gg \phi^{(2)}(d_0)/\lambda_c^2$, the two
terms in $A_2$ would not be comparable. Then the effects of
compressibility could become significant.


\end{document}